\documentstyle[prl,aps,epsf]{revtex}

\begin{document}

\advance\textheight by 0.2in

\draft
\twocolumn[\hsize\textwidth\columnwidth\hsize\csname@twocolumnfalse
\endcsname  

\title{Correlation Lengths in the Vortex Line Liquid of a High 
$T_c$ Superconductor}

\author{Peter Olsson$^1$ and S. Teitel$^2$}

\address{$^1$Department of Theoretical Physics, Ume{\aa} University, 
901 87 Ume{\aa}, Sweden}
\address{$^2$Department of Physics and Astronomy, University of Rochester, 
Rochester, NY 14627}

\date{\today}	

\maketitle

\begin{abstract}
	We use the three dimensional uniformly frustrated XY model, as
a model for a high temperature superconductor in an applied magnetic
field, to explicitly measure the longitudinal correlation length $\xi_z$
in the vortex line liquid phase.  We determine the scaling of $\xi_z$
with magnetic field and system anisotropy close to the vortex 
lattice melting transition.  We apply our results to determine
the extent of longitudinal correlations in YBCO just above melting.
\end{abstract}


]

It is now generally accepted that thermal fluctuations in the
high $T_c$ superconductors lead, for a clean sample in the mixed state, 
to a first order melting of the vortex line lattice into a vortex line liquid.
The properties of this vortex line liquid have been the subject of
considerable investigation.
An early theory by Feigel'man and co-workers \cite{R1} proposed 
that longitudinal (parallel to the applied field $H$) superconducting coherence 
could still persist above melting.  Flux transformer experiments in 
heavily $twinned$ YBCO single crystals \cite{R2} suggested 
support for this conclusion, 
as did early numerical simulations \cite{R3.1,R3.2} of the 
frustrated three dimensional (3D) XY model.  However more recent 
experiments on $untwinned$ YBCO single crystals by 
L\'{o}pez {\it et al}. \cite{R4} found longitudinal 
coherence to vanish simultaneously with melting, as have recent, more 
extensive, XY simulations by Hu {\it et al}. and by Nguyen and 
Sudb{\o} \cite{R5}.  However it remains an important open question 
just how 
large the finite longitudinal correlations can become just above melting.
Simulations by Nordborg and Blatter  \cite{R6} within the 
``two dimensional (2D) boson'' approximation, as well as general
theoretical considerations \cite{R7}, predict a correlation
length $\xi_z\sim\gamma^{-1} a_{\rm v}$, where 
$\gamma\equiv\lambda_z/\lambda_\perp$
is the anisotropy ratio and $a_{\rm v}=\sqrt{\phi_0/B}$ is the average
spacing between vortex lines.  However analyses of
experiments on untwinned single crystal YBCO
by Righi {\it et al}. \cite{R8} and by Moore \cite{R9}
have suggested that longitudinal correlations may be of the 
surprisingly larger micron scale.

To investigate this issue, we carry out extensive new simulations
of the frustrated 3D XY model for different values of applied
flux density $f$ and anisotropy $\eta$, explicitly measuring the longitudinal 
correlation length $\xi_z$ as determined by several different criteria.
We find a good scaling of $\xi_z$ with $f$ and $\eta$ in the continuum 
limit, allowing us to estimate $\xi_z(T_c)$ in real YBCO single crystal
samples. We find that longitudinal correlations at melting are
enhanced with respect to the 2D boson approximation, but not 
dramatically so.  
We also address several additional questions.
We show, contrary to recent claims \cite{R12}, that there is only a single
transition even in the isotropic model.  
In the very anisotropic limit $\xi_z(T_c)<d$, where a cross over to
2D behavior has been predicted \cite{R7,R15}, we find no qualitative
differences from the less anisotropic cases.  We find that
thermally excited vortex loops, which become important at low magnetic fields, 
can be described by an effective renormalization of the interaction between
field induced vortex lines, and we find no evidence for a recently
proposed transition within the vortex line liquid phase \cite{R17,R18}.

Our model is the uniformly frustrated 3D XY model \cite{R10}, 
given by the Hamiltonian
\begin{equation}
	{\cal H}[\{\theta_i\}]=-\sum_{i,\hat\mu}J_\mu\cos(\theta_i-
     \theta_{i+\hat\mu}-A_{i\mu})\enspace,
	\label{eq:H}
\end{equation}
where the sum is over the sites $i$ of a cubic grid of points with
unit basis vectors $\hat\mu=\hat x,\hat y, \hat z$. $\theta_i$ is 
the phase angle of the superconducting wavefunction on site $i$, and 
$A_{i\mu}=(2\pi/\phi_0)\int_i^{i+\hat\mu}{\bf A}\cdot d{\bf l}$
is the integral of the magnetic vector potential on the specified bond.
The unit of the grid spacing along
$\hat z$ is taken as $d$, the spacing between the weakly coupled 
CuO planes; the unit of the grid spacing in the $xy$ plane is
taken as $\xi_{\perp 0}$, the bare vortex core size in the plane.
The Hamiltonian (\ref{eq:H}) results from making the 
London approximation to the
discretized Ginzburg-Landau energy, and assuming $\lambda/a_{\rm v}\to\infty$ 
so that the
internal magnetic field ${\bf B}$ can be taken as frozen and equal to the 
uniform applied field ${\bf H}$.  
For a uniaxial anisotropic system with weak direction along 
$\hat z$, the couplings are $J_{x,y}\equiv J_\perp=\phi_0^2 
d/(16\pi^3\lambda_\perp^2)$ and $J_z=\phi_0^2 \xi_{\perp 0}^2
/(16\pi^3\lambda_z^2d)$, where $\lambda_\perp$ and $\lambda_z$ are
the penetration lengths in the respective directions.
The anisotropy is given by the parameter
\begin{equation}
	\eta\equiv\sqrt{J_\perp\over J_z}
    ={\lambda_z\over\lambda_\perp}{d\over\xi_{\perp 0}}
   \equiv\gamma{d\over\xi_{\perp 0}}
	\label{eq:eta}
\end{equation}
and the magnetic field is taken uniform along $\hat z$,
with a density of flux quanta per plaquette of the grid,
\begin{equation}
	f\equiv B\xi_{\perp 0}^2/\phi_0 = (\xi_{\perp 0}/a_{\rm v})^2\enspace .
	\label{eq:f}
\end{equation}
$f$ and $\eta$ are the two dimensionless parameters of our model.
A more complete derivation of Eq.~(\ref{eq:H}), and justification for
its use in modeling high $T_c$ materials, is given in Ref.~\cite{R3.2}.
Its advantage over the ``2D boson'' approximation is in its more
realistic vortex line interaction, and in that it allows for the
production of thermally activated vortex ring excitations, which may
be important at small $f$.

To determine the relevant transitions in the model, we simulate
Eq.~(\ref{eq:H}) with periodic boundary conditions \cite{RBC}, measuring the
standard quantities \cite{R10.1}: ($i$) the helicity moduli parallel 
and perpendicular to
the field, $\Upsilon_z$ and $\Upsilon_\perp$, which measure 
phase coherence, and ($ii$) $\Delta S({\bf K})=S({\bf K})-S(R_x[{\bf 
K}])$, where
$S({\bf k}_\perp)$ is the average intraplanar vortex structure
function, ${\bf K}$ is a reciprocal
lattice vector of the ordered vortex lattice, and $R_x$ reflects ${\bf 
K}$ through the $x$ axis; the difference is used so that $\Delta S$
vanishes in the liquid, and we average $\Delta S$ over the three 
smallest non-zero values of ${\bf K}$.  Our simulations at 
temperatures near the transition, for a lattice of size 
$L_\perp\times L_\perp\times 
L_z$, consist typically of $L_\perp^2L_z$ Monte Carlo passes 
through the entire lattice for equilibration \cite{R100}, 
followed by $2\times 10^6 - 10^7$ passes for computing averages.

\begin{figure}
\epsfxsize=7.5truecm
\epsfbox{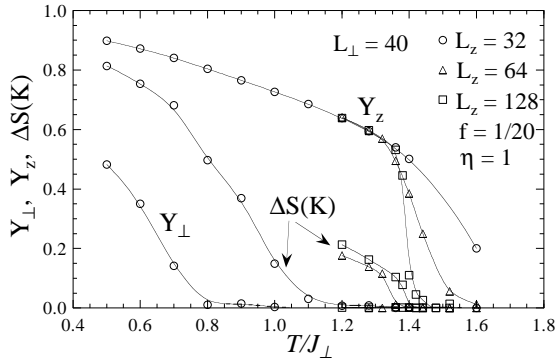}
\vspace{9pt}
\caption{Helicity moduli $\Upsilon_z$, $\Upsilon_z$ and vortex
structure order parameter $\Delta S({\bf K})$ vs. $T$, for $f=1/20$
and isotropic couplings $\eta=1$, at system sizes $L_\perp=40$ and
$L_z=32$, $64$ and $128$.  Solid lines are guides to the eye.
}
\label{fig1}
\end{figure}

An example of our results is shown in Fig.~\ref{fig1} below for the
case of isotropic couplings $\eta=1$, and $f=1/20$, for $L_\perp=40$
and several different sizes $L_z$.  If we denote the
loss of longitudinal coherence, where $\Upsilon_z$ vanishes, as
$T_c$, and the melting of the vortex lattice,  where $\Delta S$
vanishes, as $T_m$, then only
for the largest $L_z$ do we clearly observe a single transition with 
$T_m=T_c$ \cite{R11}.  The strongest finite size effect is the $increase$ in 
$T_m$ as $L_z$ increases.  Our results explain recent simulations 
by Ryu and Stroud \cite{R12} which, using smaller systems, 
continued to suggest 
$T_m<T_c$ for the isotropic model.  Note that $\Upsilon_\perp$ 
vanishes well below $T_c$, 
indicating that the vortex lattice has depinned from our numerical grid
well below its melting.

We next measure the longitudinal correlation lengths in the vortex 
line liquid, $T_c<T$, as determined three different ways.
The phase angle correlation length $\xi_z$ and the vortex correlation length
$\xi_{{\rm v}z}$ are defined by the correlation functions,
\begin{eqnarray}
	C(z) & \equiv & \langle e^{i[\theta({\bf r}_\perp,z)
    -\theta({\bf r}_\perp,0)]}\rangle\ \ \sim  e^{-z/\xi_z},\ \quad T_c<T
 \label{eq:xiz}\\
	C_{\rm v}(z) & \equiv & \langle n_z({\bf r}_\perp,z)n_z({\bf 
    r}_\perp,0)\rangle \sim e^{-z/\xi_{{\rm v}z}},\quad T_c<T
 \label{eq:xivz}
\end{eqnarray}
Here $n_z({\bf r}_\perp,z)={1\over 2\pi}
[{\bf D}\times {\bf D}\theta]\cdot\hat z$ 
is the vorticity in the $xy$ plane 
at transverse position ${\bf r}_\perp$ and height $z$ (${\bf D}$ is
the lattice difference operator).  We work in a
gauge for which $A_{iz}=0$.  
Our third length is determined by considering the wavevector dependent
helicity modulus $\Upsilon_z({\bf k}_\perp)$, which gives the linear
response in supercurrent to a perturbation in vector potential 
$A_z({\bf k}_\perp)$ \cite{R13}.  In $D=3$, dimensional analysis gives
$\Upsilon_z\sim~1/{\rm length}$.  In the vortex liquid, 
provided one is not near any critical point 
where anomalous dimensions might come into play \cite{R14},
$\Upsilon_z({\bf k}_\perp)$ must vanish as $k_\perp^2$ as 
$k_\perp\to 0$.  We therefore define the helicity correlation 
length $\xi_{\Upsilon z}$ by
\begin{equation}
	\Upsilon_z({\bf k}_\perp)\equiv c\> \xi_{\Upsilon z} 
     k_\perp^2,\qquad T_c<T
	\label{eq:xiUps}
\end{equation}
where c is a constant numerical factor, which we fix in an
{\it ad hoc} manner by requiring
$\xi_{\Upsilon z}=\xi_z$ at $T=1.2$.  

For each case we have considered, we first carefully choose 
$L_z$ sufficiently large so as to 
observe a single sharp first order melting transition;
however $L_z$ must not be $too$ large, in order that we are still able 
to cool into the vortex lattice state without getting trapped in
a supercooled liquid.  For such a value of $L_z$, we carefully
monitor the time
sequence of $\Delta S$ and determine $T_c$
as the temperature at which the system seems to be switching
equally between vortex lattice and vortex liquid states.
To accurately measure correlation lengths, we then repeat
the simulations with a larger value of $L_z\gg \xi_z(T_c)$,
cooling down to the predetermined $T_c$.  

\begin{figure}
\epsfxsize=7.5truecm
\epsfbox{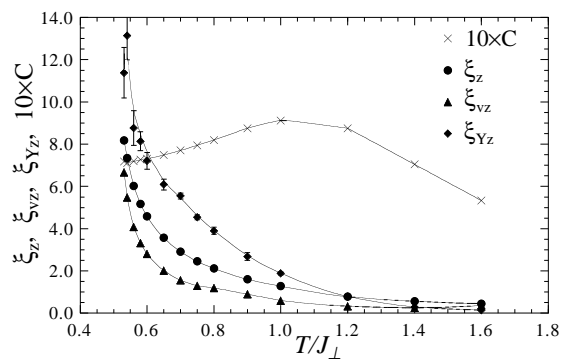}
\vspace{9pt}
\caption{Phase, vortex, and helicity correlation lengths, $\xi_z$,
$\xi_{{\rm v}z}$, and $\xi_{\Upsilon z}$ vs. $T$ for $f=1/20$ and
$\eta^2=9$, for system size $L_\perp=40$ and $L_z=128$.  Also
shown is the specific heat $C$.  Solid lines are guides to the eye.
}
\label{fig2}
\end{figure}

In Fig.~\ref{fig2} we show results for $\xi_z$, $\xi_{{\rm v}z}$ and 
$\xi_{\Upsilon z}$ vs. $T$, for the case of $f=1/20$, $\eta^2=9$, with
$L_\perp=40$ and $L_z=128$ (for $T\geq 0.8$, $L_z=64$).   
We also show the specific heat $C$,
as computed from energy fluctuations.  The peak in $C$ at ``$T_{c2}$''
is identified as the cross-over where, upon cooling, local 
superconducting order first develops \cite{R3.2}.
We see that all three lengths increase similarly as one cools towards 
$T_c$.   No noticeable feature is seen near $T_{c2}$.    
$\xi_z$ is slightly larger than $\xi_{{\rm v}z}$ by a factor of about $1.3$.
The numerical factor of Eq.~(\ref{eq:xiUps}) is found to be $c=74$.
In determining $\xi_z$ and $\xi_{{\bf v}z}$ from Eqs.~(\ref{eq:xiz}) 
and (\ref{eq:xivz}), we fit our data self consistently within the
range $\xi_z<z<L_z/3$, averaging over the position ${\bf r}_\perp$.
In determining $\xi_{\Upsilon z}$, we fit Eq.~(\ref{eq:xiUps})
to the two smallest non-zero values of $k_\perp$, since we found
that $\Upsilon_z({\bf k}_\perp)$ quickly saturated to a constant as
$k_\perp$ increased.  This unfortunately limits the accuracy
with which we can determine $\xi_{\Upsilon z}$.
Some examples of our fits for $\xi_z$ and $\xi_{\Upsilon z}$ are shown
in Fig.~\ref{fig3} below.
In the remainder of this work we now focus on
the phase correlation length $\xi_z$.

\begin{figure}
\epsfxsize=7.5truecm
\epsfbox{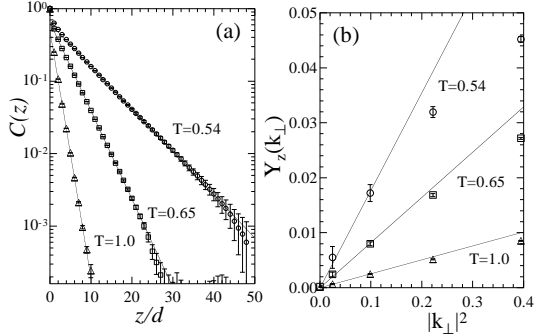}
\vspace{9pt}
\caption{$a$) Phase correlation $C(z)$ vs. $z$, and $b$) helicity
$\Upsilon_z({\bf k}_\perp)$ vs. $k_\perp$, for several different
values of $T$ for the parameters of Fig.~\ref{fig2}.  
The solid lines are fits to 
Eqs.~(\ref{eq:xiz}) and (\ref{eq:xiUps}) that determine $\xi_z$
and $\xi_{\Upsilon z}$.
}
\label{fig3}
\end{figure}

\begin{figure}
\epsfxsize=7.5truecm
\epsfbox{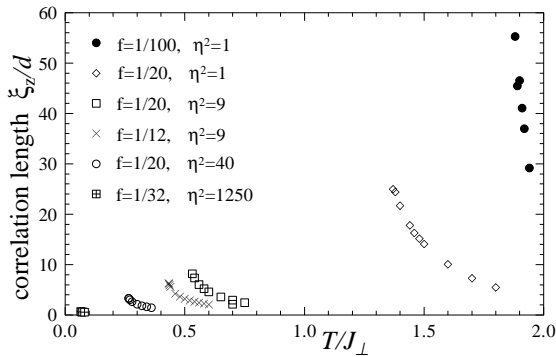}
\vspace{9pt}
\caption{Phase correlation length $\xi_z$ vs. $T$ for parameter
values $f=1/20$ and $\eta^2=1$, $9$, $40$ (system sizes are $L_\perp=40$
and $L_z=192$, $128$, $32$ respectively); $f=1/12$ and $\eta^2=9$
(system size is $L_\perp=24$ and $L_z=64$);
$f=1/32$ and $\eta^2=1250$ (system size is $L_\perp=64$ and $L_z=8$);
and $f=1/100$ and $\eta^2=1$ (system size is $L_\perp=100$ and $L_z=256$).
}
\label{fig4}
\end{figure}

In Fig.~\ref{fig4} we show our results for $\xi_z$ vs. $T$, for 
several different values of the parameters $f$ and $\eta$.

\begin{figure}
\epsfxsize=7.5truecm
\epsfbox{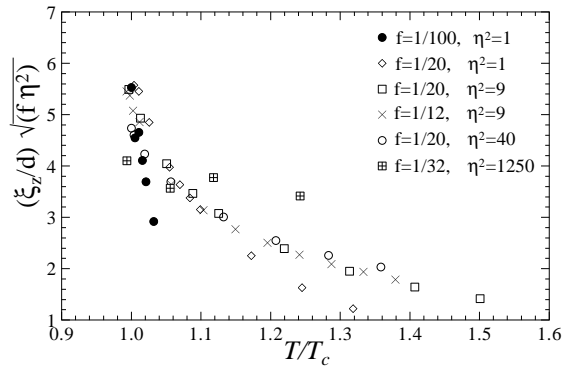}
\vspace{9pt}
\caption{Data of Fig.~\ref{fig4} replotted as $(\xi_z/d)\sqrt{f\eta^2}$
vs. $T/T_c$.
}
\label{fig5}
\end{figure}

We can now argue as follows for the dependence of $\xi_z$
on $f$ and $\eta$.  For small values of $f$, such that $\xi_{\perp 
0}\ll a_{\rm v}$, we expect that $\xi_z$ should be independent
of the vortex core size $\xi_{\perp 0}$.  From 
Eqs.~(\ref{eq:eta}-\ref{eq:f}) we see that the only combination
of $f$ and $\eta$ that is independent of $\xi_{\perp 0}$ is $f\eta^2$.
Furthermore, for large $\xi_z$ we expect our discretizing grid
to become a reasonable approximation of the continuum and so 
$\xi_z$ should be independent of the layer spacing $d$.  We therefore
expect that the dimensionless $\xi_z/d$ should scale as $1/d$, and
so we conclude $\xi_z/d\sim 1/\sqrt{f\eta^2}$.  In Fig.~\ref{fig5} below
we replot the results of Fig.~\ref{fig4} as $(\xi_z/d)\sqrt{f\eta^2}$
vs. $T/T_c$.  We see that as $T\to T_c$, most of the data collapse
to a single curve.  
Deviations from this curve represent situations when
either $\xi_z/d$ is small $\sim O(1)$, or when $T$ is
sufficiently large (approaching $T_{c2}$)
that thermally excited vortex rings start
to dominate the total vorticity of the system.  In the
first case, the discreteness of our grid spacing $d$ clearly
becomes an important length scale.  In the second case, as the
density of thermal rings is determined by the vortex
core energy, and hence by the core sizes $d$ and $\xi_{\perp 0}$,
again the discreteness of our grid becomes evident.
Such deviations thus occur for all cases at sufficiently high $T$, and
also for the case $f=1/32$, $\eta^2=1250$ at all $T$.  This latter
case was specifically chosen so that, according to continuum expressions,
one would expect $\xi_z(T_c)<d$ and so to be in the so-called ``2D'' limit
of very weakly coupled layers \cite{R7,R15}.  We see that for 
this case $\xi_z(T_c)/d$
lies below the other data, indicating an even smaller correlation
length than one would expect in a continuum.  However 
we otherwise found no anomalous behavior for this case: there remained
only a single transition where longitudinal coherence and vortex 
lattice order vanished simultaneously.

Except for the ``2D'' case discussed above, our other data, when 
appropriately scaled as in Fig.~\ref{fig5}, all coincide at $T_c$.
We therefore conclude that, for these cases, our model is well
approximating continuum behavior.  From the specific numerical value
of $(\xi_z/d)\sqrt{f\eta^2}$ at $T_c$ in Fig.~\ref{fig5} we can
therefore conclude that in a uniaxial anisotropic superconductor in
the ``3D'' continuum limit,
\begin{equation}
	\xi_z(T_c) \simeq 5.5 d/\sqrt{f\eta^2}= 5.5 \gamma^{-1} a_{\rm v}\enspace ,
	\label{eq:result}
\end{equation}
where the second equality follows from Eqs.~(\ref{eq:eta}-\ref{eq:f}),
with $\gamma\equiv\lambda_z/\lambda_\perp$ 
the anisotropy, and $a_{\rm v}$ the average
spacing between vortex lines.  Applying this result to YBCO, for which
$\gamma\sim 7$, we conclude that $\xi_z(T_c)\simeq 0.86 a_{\rm v}$, or
$\xi_z\simeq 0.023\mu$ for a field of $B\simeq H=4$T.

Our result above may be compared with that of recent ``2D boson'' 
simulations of Nordborg and Blatter \cite{R6}, which yielded $\xi_{{\rm 
v}z}(T_c)=1.7 \gamma^{-1}a_{\rm v}$.  If we take from Fig.~\ref{fig2}
that $\xi_z(T_c)\simeq 1.3\xi_{{\rm v}z}(T_c)$, then
the more realistic vortex
line interaction of the XY model gives roughly a $2.5$ fold increase
in $\xi_z(T_c)$ over the boson model.  
This remains, however, well below the micron scale.

Except for the case $f=1/100$, 
our numerical results are all in the limit of sufficiently
large $f$, such that $T_c(f,\eta)$ lies well below the zero field 
critical point $T_c(0,\eta)$.  For these cases, therefore, thermally 
excited vortex rings are $not$ playing any significant 
role at our melting transitions.  
One way to see this is to note that for these cases, the melting
temperatures $T_c(f,\eta)$ obey quite well the expectation of
the Lindemann criterion (which ignores thermal rings),
$T_c/J_\perp\propto 1/\sqrt{f\eta^2}$ \cite{R3.2}.  To see this,
note our result above that $\xi_z(T_c)/d \propto 1/\sqrt{f\eta^2}$,
and hence, if the Lindemann criterion holds, we expect
$[\xi_z(T_c)/d]/[T_c/J_\perp]$ to be a constant.  In Fig.~\ref{fig4}
we see that the loci of points $([\xi_z(T_c)/d],[T_c/J_\perp])$ do 
indeed lie on quite close to a straight line intersecting the origin.

The case $f=1/100$, however, clearly lies off this line.  This is as
expected: as $f$ decreases and $T_c(f,\eta)$ increases, one eventually 
enters the critical region of the $f=0$ transition, where thermally excited
rings play a significant role in renormalizing the effective 
interactions between the magnetic field induced vortex lines, 
and suppress the melting transition below the value predicted by
the ``bare'' Lindemann criterion, so that $\lim_{f\to 0} 
T_c(f,\eta)=T_c(0,\eta)$.  In this
case, our argument that $\xi_z(T_c)$ should be independent of
$\xi_{\perp 0}$ becomes less obvious.  Nevertheless, we see in 
Fig.~\ref{fig5} that the data for $f=1/100$ agrees quite well
with our scaling assumption in the near vicinity of $T_c$.
We therefore conclude that the main effect of the thermally excited 
rings at melting is indeed adequately described by a renormalization of
vortex line couplings \cite{R16}, and so
our result of Eq.~(\ref{eq:result}) will continue
to hold in the low field region, although with a possible 
renormalization of the anisotropy parameter $\gamma$.

Recently, Te\v{s}anovi\'{c} \cite{R17} has argued that there may still be
a singular vortex ring blowup transition at low fields, within the 
normal vortex line liquid.  Recent XY simulations by Nguyen and Sudb{\o} 
\cite{R18} have
claimed to identify this transition in terms of a sharp percolation 
transition of  transverse vortex paths, which takes place in the
vicinity of the peak in the specific heat.  They find that,
within the XY model, this
percolation transition is more clearly distinct from the melting 
transition at
$larger$ rather than $smaller$ fields.  One of Te\v{s}anovi\'{c}'s
predictions is that the length $\xi_{\Upsilon z}$ will have a 
discontinuous decrease at this transition.  This prediction has been
one of our motivations in computing $\xi_{\Upsilon z}$.
In Fig.~\ref{fig2} we find no clear evidence for such behavior
in $\xi_{\Upsilon z}$.
It therefore remains unclear, if such a percolation
or ring blowup transition does exists, whether it has any noticeable
effect on thermodynamically measurable quantities.

We wish to thank Z. Te\v{s}anovi\'{c} for many stimulating 
discussions.  This work has been supported by U.S. 
DOE grant DE-FG02-89ER14017,
by Swedish Natural Science Research Council Contract No. E-EG 10376-305,
and by the resources of the Swedish High Performance Computing Center 
North (HPC2N).


\begin{thebibliography}{99}
\bibitem{RBC}We have also tried ``fluctuating
   twist'' boundary conditions, in an attempt to minimize finite
   size effects (see Olsson in Ref.~\protect\cite{R13}, and Ref.~\protect
   \cite{R14}).  However, for the large values of $L_z$
   used here, there was a negligible difference between results from
   the different boundary conditions.
\bibitem{R100}For $f=1/100$, we used $L_\perp^2L_z/15$ passes
   for equilibration.
\end{thebibliography}
\end{document}